\documentclass[twocolumn,prb]{revtex4}
\usepackage{amsfonts}
\usepackage[T1]{fontenc}
\usepackage{amsmath,amsbsy,amssymb,graphicx}
\usepackage{times}

\let\mathbf=\boldsymbol

\begin{document}

\title{ Loop-nodal and Point-nodal Semimetals in Three-dimensional Honeycomb
Lattices}
\author{Motohiko Ezawa}
\affiliation{Department of Applied Physics, University of Tokyo, Hongo 7-3-1, 113-8656,
Japan}

\begin{abstract}
Honeycomb structure has a natural extension to the three dimensions. Simple
examples are hyperhoneycomb and stripy-honeycomb lattices, which are
realized in $\beta $-Li$_{2}$IrO$_{3}$ and $\gamma $-Li$_{2}$IrO$_{3}$,
respectively. We propose a wide class of three-dimensional (3D) honeycomb
lattices which are loop-nodal semimetals. Their edge states have intriguing
properties similar to the two-dimensional honeycomb lattice in spite of
dimensional difference. Partial flat bands emerge at the zigzag or beard
edge of the 3D honeycomb lattice, whose boundary is given by the Fermi loop
in the bulk spectrum. Analytic solutions are explicitly constructed for
them. On the other hand, perfect flat bands emerge in the zigzag-beard edge
or when the anisotropy is large. All these 3D honeycomb lattices become
strong topological insulators with the inclusion of the spin-orbit
interaction. Furthermore, point-nodal semimetals may be realized in the
presence of both the antiferromagnetic order and the spin-orbit interaction.
\end{abstract}

\maketitle

Honeycomb lattice is materialized naturally in graphene and in related
materials, which presents one of the most active fields of condensed matter
physics. The Fermi surface is zero dimensional, given by the $K$ and $K^{\prime }$ points, 
though in general the dimension of the Fermi surface is 
$D-1$ for the $D$-dimensional system. It may be called a point-nodal
semimetal with the Dirac cone. A nanoribbon made of honeycomb lattice has
interesting properties such as zero-energy flat bands connecting the $K$ and 
$K^{\prime }$ points\cite{Klein,Fujita,EzawaGNR}. Perfect flat bands are
generated when the $K$ and $K^{\prime }$ points are shifted and merged by
introducing anisotropy in the hopping\cite{Wunsch,Pere,Monta}, which is
realized in phosphorene\cite{Phos}. A natural question is whether there are
similar properties in the three dimensions.

Honeycomb structure has a natural extension to the three dimensions.
Examples are hyperhoneycomb\cite{Takagi,Biffin,Modic} and stripy-honeycomb\cite{Modic} lattices, 
being realized in $\beta $-Li$_{2}$IrO$_{3}$ and $\gamma $-Li$_{2}$IrO$_{3}$, respectively. 
A series of three-dimensional (3D)
honeycomb lattices named harmonic honeycomb lattices\cite{Modic} have also
been proposed. They attracts much attention\cite{Lee,LeeAF,Mullen,Nasu,Kimchi,LeeO,Nasu2,Hermanns} recently. It has been
argued\cite{Lee,Mullen} that the hyperhoneycomb lattice is a loop-nodal
semimetal where the Fermi surface forms a loop (which we call a Fermi loop).
Furthermore, the system becomes a topological insulator by introducing a
spin-orbit interaction\cite{Lee} (SOI). Various antiferromagnetic order is
reported in the hyperhoneycomb lattice\cite{LeeAF,LeeO,Kimchi,Kimchi2}.
Similar Fermi loops have been predicted in other systems\cite{Philip,Xie,Yu,Kim} based on first-principles calculation.

In this Letter, we propose a wide class of 3D honeycomb lattices which are
loop-nodal semimetals. We first analyze the edge states of nanofilms made of
them. Flat-band edge states emerge at the zigzag or beard edge termination,
which are reminiscence of the edge states of the honeycomb system. We derive
an analytic form of the wave function by the recursion method. The boundary
of the zero-energy states is given by the Fermi loop in the bulk spectrum.
It is shown that the perfect flat band is generated over the whole region of
the Brillouin zone by two methods: One is terminating the sample with the
zigzag and beard edges; The other is increasing the anisotropy, where the
Fermi loop shrinks and disappears. All these 3D honeycomb lattices become
strong topological insulators in the presence of the spin-orbit interaction.
Furthermore, a point-nodal semimetal may be generated together with a Dirac
cone in the additional presence of the antiferromagnetic order.

\begin{figure}[t]
\centerline{\includegraphics[width=0.5\textwidth]{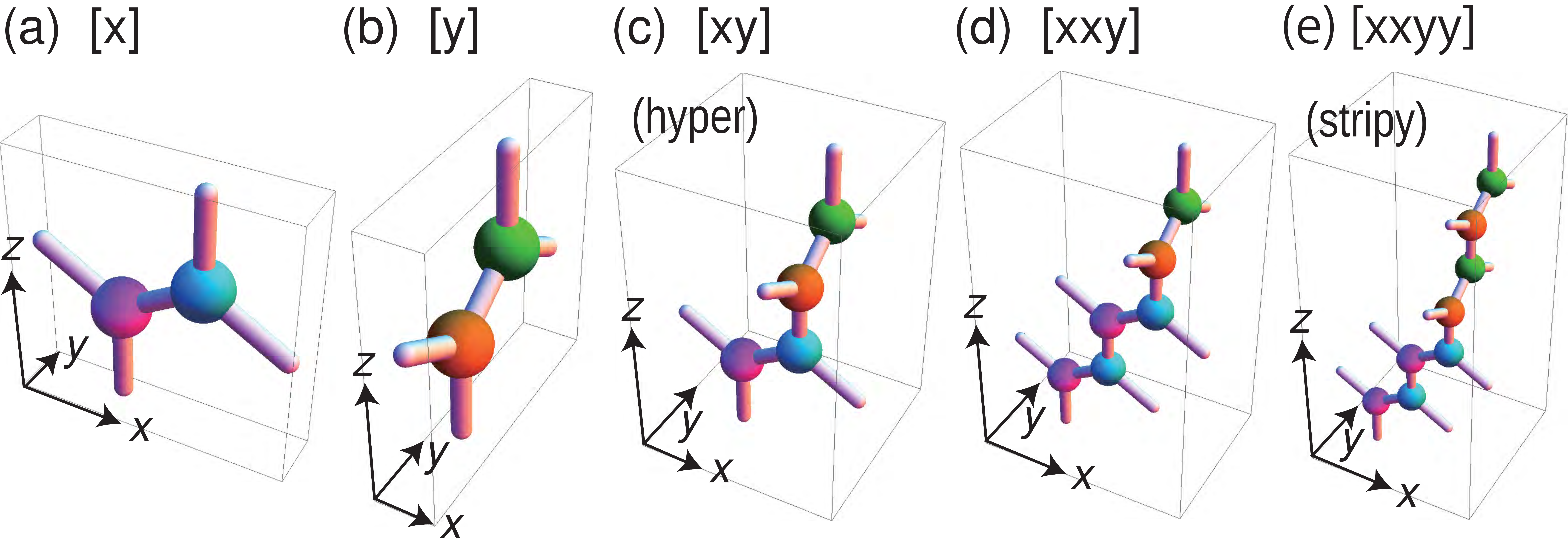}}
\caption{Unit cells of various honeycomb lattices. The unit cell $[\protect\alpha _{1}]$ 
of the honeycomb lattice placed on (a) the $xz$-plane and (b)
the $yz$-plane. We use these building blocks to construct 3D honeycomb
lattices. (c,d,e) Typical examples of unit cells $[\protect\alpha _{1}\protect\alpha _{2}\cdots \protect\alpha _{N}]$ 
made by sewing the above building blocks. }
\label{FigYJunc}
\end{figure}

\textit{Lattice structure and model:} Honeycomb lattice is a bipartite
system. The unit cell contains two vertices (atoms) and five links (bonds)
making the angle $2\pi /3$ between the neighboring ones with certain links
being identified on one plane. We prepare two sets placed on the $xz$ plane
and the $yz$ plane, and refer to them as the building blocks $[x]$ and $[y]$, respectively: 
See illustration in Fig.\ref{FigYJunc}(a) and (b). We
propose a class of 3D honeycomb lattices by sewing these two building blocks
in such a way that all atoms in one unit cell are connected with a single
path tending to the $z$ direction [Fig.\ref{FigYJunc}(c), (d) and (e)].

Let us first consider the unit cell containing $4$ atoms. It is uniquely
given by $[xy]$ if we start with $[x]$. Note that $[xx]$ is equivalent to $[x]$. 
In general, the unit cell containing $N$ building blocks is
represented by $[\alpha _{1}\alpha _{2}\cdots \alpha _{N}]$, where $\alpha
_{n}=x$ or $y$. There exist the cyclic symmetry; namely, two unit cells 
$[\alpha _{1}\alpha _{2}\cdots \alpha _{N-1}\alpha _{N}]$ and 
$[\alpha_{2}\alpha _{3}\cdots \alpha _{N}\alpha _{1}]$ are equivalent. 
Furthermore, $[\alpha _{1}\alpha _{2}\cdots \alpha _{N}]$ and 
$[\bar{\alpha}_{1}\bar{\alpha}_{2}\cdots \bar{\alpha}_{N}]$ are equivalent, where $\bar{\alpha}_{n}=y$ $(x)$ 
if $\alpha _{n}=x$ $(y)$. The simplest 3D honeycomb lattice is
generated by the unit cell $[xy]$, which has been named the hyperhoneycomb
lattice. The next simplest one is $[xxy]$. Then, we have $[xxyy]$, which
generates the stripy-honeycomb lattice. The type of lattices with the unit
cell $[x\cdots xy\cdots y]$ in sequential order of $x$ and $\ y$ has been
named the harmonic honeycomb lattice when the numbers of $x$'s and $y$'s are
equal.

\textit{Loop-nodal semimetal:} We consider a free-electron system hopping on
the 3D honeycomb lattice whose unit is $[\alpha _{1}\alpha _{2}\cdots \alpha
_{N}]$. The Hamiltonian is given $H=\sum_{\langle i,j\rangle
}t_{ij}c_{i}^{\dagger }c_{j},$ where $t_{ij}=t_{z}$ for the nearest neighbor
hopping along the $z$ axis and $t_{ij}=t_{xy}$ for the other
nearest-neighbor hopping and $c_{i}$ ($c_{i}^{\dagger }$) is the
annihilation (creation) operator of the electron at the site $i$. In the
momentum representation it is given by the $2N\times 2N$ matrix\cite{Lee}, 
$H_{2N}=\sum_{\mathbf{k}}c^{\dagger }(\mathbf{k})\hat{H}_{2N}(\mathbf{k})c(\mathbf{k}),$ 
with $H_{2n-1,2n}=f_{\alpha _{n}}$, $H_{2n,2n-1}=f_{\alpha
_{n}}^{\ast }$, $H_{2n,2n+1}=f_{z}$, $H_{2n+1,2n}=f_{z}^{\ast }$, 
$H_{2N,1}=f_{z}$, $H_{1,2N}=f_{z}^{\ast }$ and all other elements being zero,
where
\begin{equation}
f_{\alpha }=2t_{xy}e^{i\frac{k_{z}}{2}}\cos \frac{\sqrt{3}}{2}k_{\alpha
},\qquad f_{z}=t_{z}e^{ik_{z}},
\end{equation}
with $\alpha =x,y$.

\begin{figure}[t]
\centerline{\includegraphics[width=0.42\textwidth]{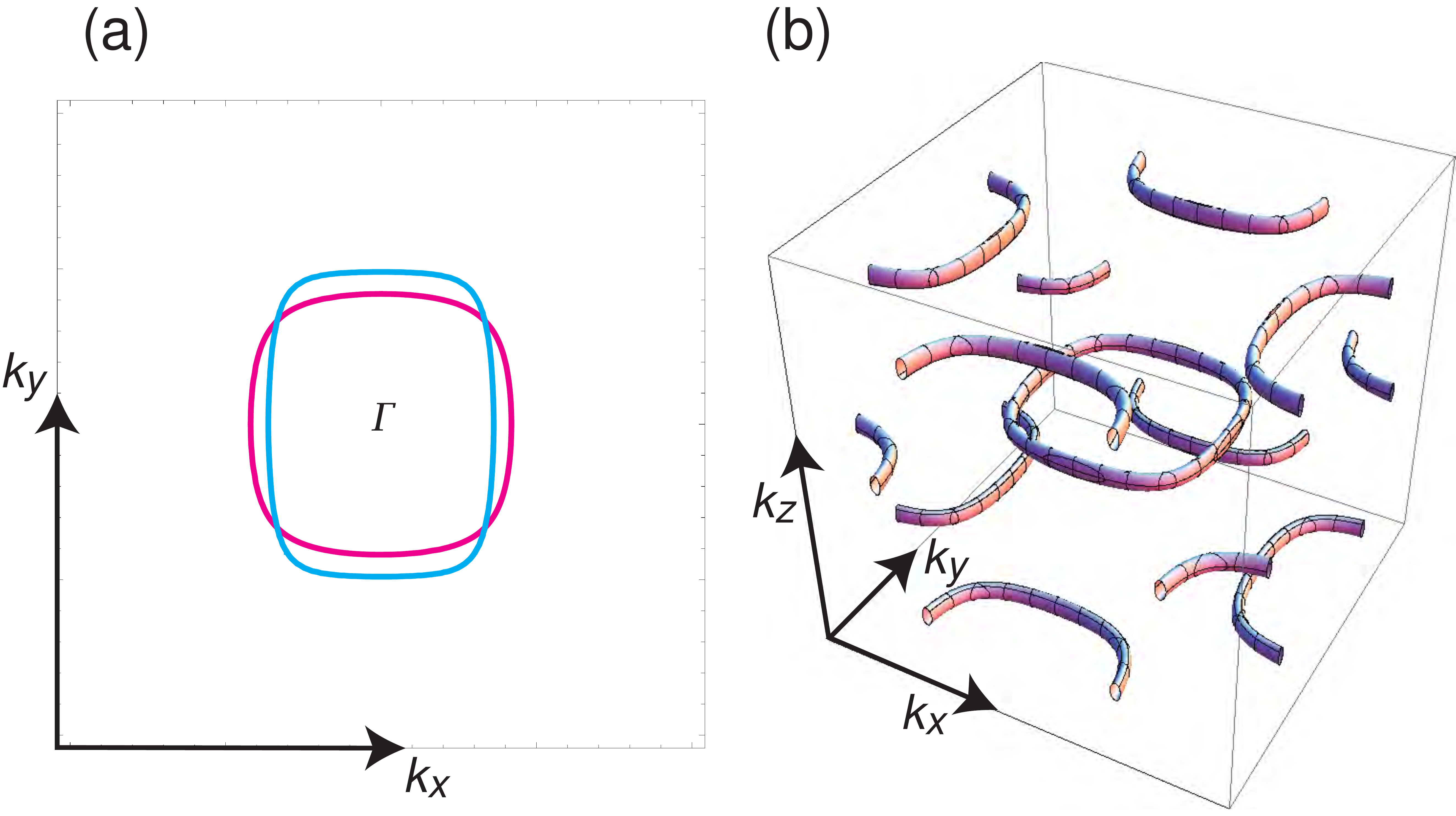}}
\caption{(a) Fermi loops with $N_{x}=N_{y}$ (magenta) and $N_{x}=4N_{y}$
(cyan), which appear in the $k_{x}k_{y}$ plane at $k_{z}=0$. We have set $t_{z}=t_{xy}=1$. 
(b) Bird's eye's view of the contour plot of the
equi-energy surface with $N_{x}=N_{y}$ for $E=0.1t$ with $t=t_{xy}=t_{z}$ in
a part of the extended Brillouin zone. }
\label{FigContour}
\end{figure}

The energy spectrum is determined by $\det \left( \lambda I-H\right) =0$.
Especially, the zero-energy states are solutions of $\det H=0$, which is
calculated as
\begin{equation}
\det H=|f_{z}^{\ast N}-\prod\limits_{n=1}^{N}f_{\alpha _{n}}|^{2}.
\end{equation}
The solution is given by $k_{z}=0$ and
\begin{equation}
\cos ^{N_{x}}\frac{\sqrt{3}}{2}k_{x}\cos ^{N_{y}}\frac{\sqrt{3}}{2}k_{y}=(t_{z}/2t_{xy})^{N_{x}+N_{y}},  \label{CondiLoop}
\end{equation}
where $N_{x}$ $(N_{y})$ is the number of $x$'s ($y$'s) in $[\alpha
_{1}\alpha _{2}\cdots \alpha _{N}]$. It represents a loop in the momentum
space [Fig.\ref{FigContour}(a)]. Hence, all 3D honeycomb lattices in this
class are loop-nodal semimetals. It follows that the Fermi surface exists
only for $\left\vert t_{xy}/t_{z}\right\vert >1/2$, as agrees with the
previous result for the case of the hyperhoneycomb lattice\cite{Lee,Mullen}
with $t_{xy}=t_{z}$.

We may illustrate the equi-energy surface of the hyperhoneycomb lattice in
the momentum space for a typical value of $E$ in Fig.\ref{FigContour}(b). It
shows a toroidal Fermi surface around the $\Gamma $ point ($\mathbf{k}=\mathbf{0}$). 
It shrinks to a 1D circle forming a Fermi loop at the half
filling with $E=0$.

\textit{Analytic wave function of the zero-energy edge states:} We first
analyze the edge states of the 3D honeycomb lattice, whose edge is taken at $z=0$. 
The momentum $k_{x}$ and $k_{y}$ remain to be good quantum numbers.
The wave function is constructed analytically with the aid of the recursive
method as in the case of the honeycomb system\cite{Kohmoto}. We define the
wave function $\psi _{i}$ for the $i$-th sites counting from the bottom of
the sample.

The wave functions of the zero-energy states must satisfy $H\psi =0$, which
is explicitly given by
\begin{equation}
S_{\pm }\equiv t_{xy}(1+e^{i\sqrt{3}k_{\alpha _{n}}})\psi _{2n\pm
1}+t_{z}\psi _{2n\mp 1}=0,
\end{equation}
for the odd sites of the zigzag ($S_{+}$) and beard ($S_{-}$) edges, 
and $\psi _{2n}=0$ for the even sites. The wave function can be solved
recursively as
\begin{equation}
\psi _{2n+1}=\left( -\frac{t_{xy}}{t_{z}}\right) ^{\pm
n}\prod\limits_{j=1}^{n}(1+e^{i\sqrt{3}k_{\alpha _{j}}})^{\pm 1}\psi _{1}
\label{WaveFunct}
\end{equation}
for the zigzag ($+$) and beard ($-$) edges.

For the zero-energy state, the wave function must take the maximum value at
the outermost edge sites and its absolute value must decrease as the site
index $i$ increases. Otherwise, the wave function diverges inside the bulk
and we cannot normalize it. This condition is given by
\begin{equation}
\cos ^{N_{x}}\frac{\sqrt{3}}{2}k_{x}\cos ^{N_{y}}\frac{\sqrt{3}}{2}k_{y}\lessgtr (t_{z}/2t_{xy})^{N_{x}+N_{y}}  \label{BBCon}
\end{equation}
for the zigzag edge ($<$) and for the beard edge ($>$). Accordingly, the
zero-energy states emerge in the region whose boundary is the Fermi loop (\ref{CondiLoop}) in the bulk spectrum.

\begin{figure*}[t]
\centerline{\includegraphics[width=0.99\textwidth]{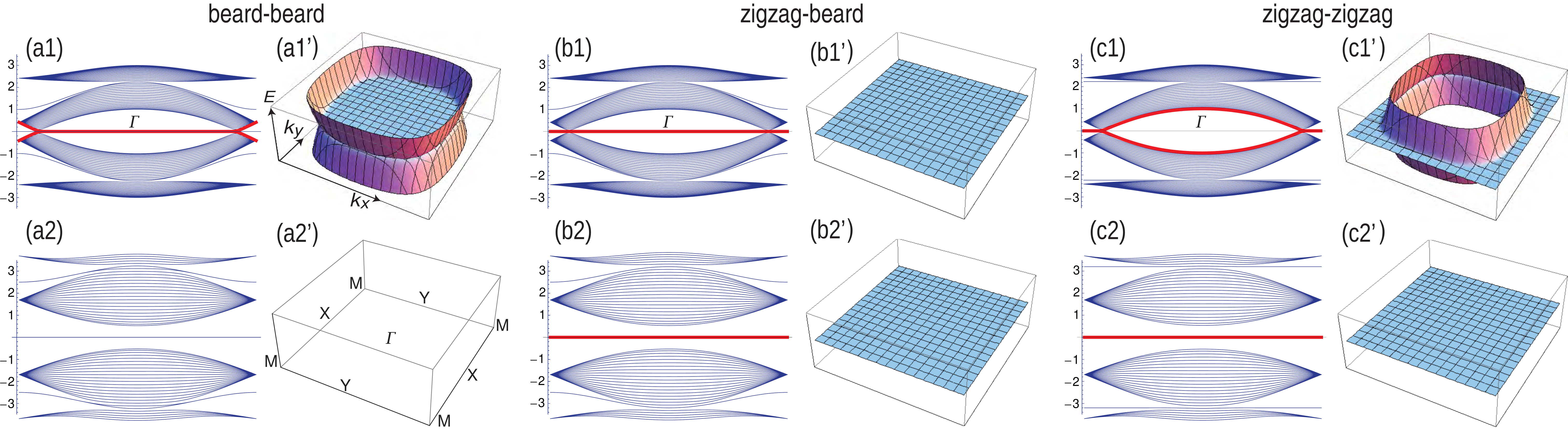}}
\caption{Band structures of nanofilms made of isotropic and anisotropic
hyperhoneycomb lattices. Edge states are marked in red. (a1),(c1) Pertial
flat bands appear for $|t_{z}|<2|t_{xy}|$ in the inner (outer) region of the
Fermi loop when the both edges are terminated by the beard (zigzag) edge.
(b) The perfect flat bands appear over the whole Brillouin zone irrespective
to the value of $t_{z}$ and $t_{xy}$ when the edges are terminated by the
zigzag and beard edges. (a2),(b2),(c2) The perfect flat bands appear for $|t_{z}|>2|t_{xy}|$ 
irrespective to the type of edge termination. We have set 
$t_{xy}=1,V=0,\protect\lambda _{xy}=\protect\lambda _{z}=0$. We have also
set $t_{z}=1$ for (a1), (b1), (c1) (isotropic), and $t_{z}=2.5$ for (a2),
(b2), (c2).}
\label{FigCompress}
\end{figure*}

It follows from (\ref{WaveFunct}) that $\psi _{1}=1$ and $\psi _{i}=0$ for 
$i\geq 2$ at the M point ($k_{x}=k_{y}=\pi /\sqrt{3}$) for the zigzag edge,
which represents the perfectly localized state corresponding to the
perfectly localized state at the zigzag edge of the honeycomb lattice.
Contrary to the zigzag edge states, there is no perfectly localized edge
state in the beard edge states.

We proceed to consider a nanofilm where the width along the $z$ direction is
finite. Both of the zigzag and beard edge terminations are possible
depending on the position of the edges. For definiteness, we show the band
structure of a nanofilm made of the hyperhoneycomb lattice in Fig.\ref{FigCompress}. 
Flat bands emerge in the band structure, which are
reminiscence of flat bands in the zigzag and beard edges in the honeycomb
lattice. When the one edge is terminated by the zigzag edge and the other
edge is terminated by the beard edge, the perfect flat bands emerge over the
whole Brillouin zone [Fig.\ref{FigCompress}(b)].

\textit{Anisotropic 3D honeycomb lattice:} We next investigate how the edge
states are modified by changing the transfer energy $t_{z}$ and $t_{xy}$.
The ratio $t_{xy}/t_{z}$ can be tuned by applying uni-axial pressure. We
show the band structure with (a) the zigzag-zigzag edges, (b) the
zigzag-beard edges, and (c) the beard-beard edges in Fig.\ref{FigCompress}
for typical values of $t_{xy}$ and $t_{z}$.

The flat band region shrinks as the $t_{z}/t_{xy}$ increases and disappears
for $|t_{z}/t_{xy}|>2$ for the beard-beard edge [Fig.\ref{FigCompress}(a)].
It is natural since there is no solution of $k_{x}$ and $k_{y}$ for $|t_{z}/t_{xy}|>2$ 
in eq.(\ref{BBCon}) with $<$. On the contrary, the flat
band region expands as $t_{z}/t_{xy}$ increases and the whole region of the
Brillouin zone becomes the perfect flat band for $|t_{z}/t_{xy}|>2$ for the
zigzag-zigzag edge [Fig.\ref{FigCompress}(b)]. This can be understood that
the condition (\ref{BBCon}) with $>$ is satisfied irrespective of the values
of $k_{x}$ and $k_{y}$ for $|t_{z}/t_{xy}|>2$.

These feature are reminiscence of the anisotropic 2D honeycomb lattice\cite{Wunsch,Pere,Monta}, 
where there are two different transfer energies $t_{1}$
and $t_{2}$. There are $K$ and $K^{\prime }$ points in the honeycomb lattice
when $t_{1}=t_{2}$. These $K$ and $K^{\prime }$ points move by changing the
ratio of the transfer energies $t_{2}/t_{1}$. When $\left\vert
t_{2}/t_{1}\right\vert >2$ they merges resulting in an insulator, where the
band dispersion is highly anisotropic. Phosphorene, monolayer black
phosphorus, is understood in this picture\cite{Phos}.

\begin{figure*}[t]
\centerline{\includegraphics[width=0.94\textwidth]{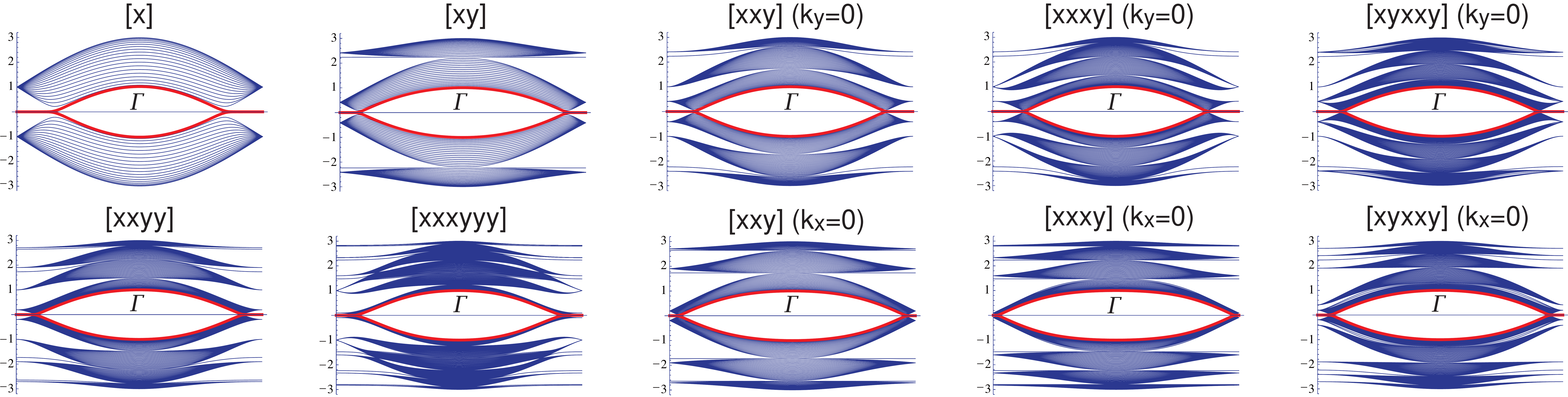}}
\caption{Band structures of nanofilms made of typical 3D-honeycomb lattices
indexed by $[\protect\alpha _{1}\protect\alpha _{2}\cdots \protect\alpha _{N}]$. 
Edge states are marked in red. All nanofilms have similar band
structures near the Fermi level. }
\label{FigRibbon}
\end{figure*}

We have so far used the instance of the hyperhoneycomb lattices for
illustration. Here we show the band structures of nanofilms made of various
3D honeycomb lattices in Fig.\ref{FigRibbon}. We find that the flat-band
zero-energy edge states emerge in the same region of the hyperhoneycomb
lattice although the high-energy band structure is different. We note that
the band structure along the $k_{x}$ and $k_{y}$ directions are inequivalent
for the 3D honeycomb lattice with $N_{x}\neq N_{y}$.

\begin{figure}[t]
\centerline{\includegraphics[width=0.5\textwidth]{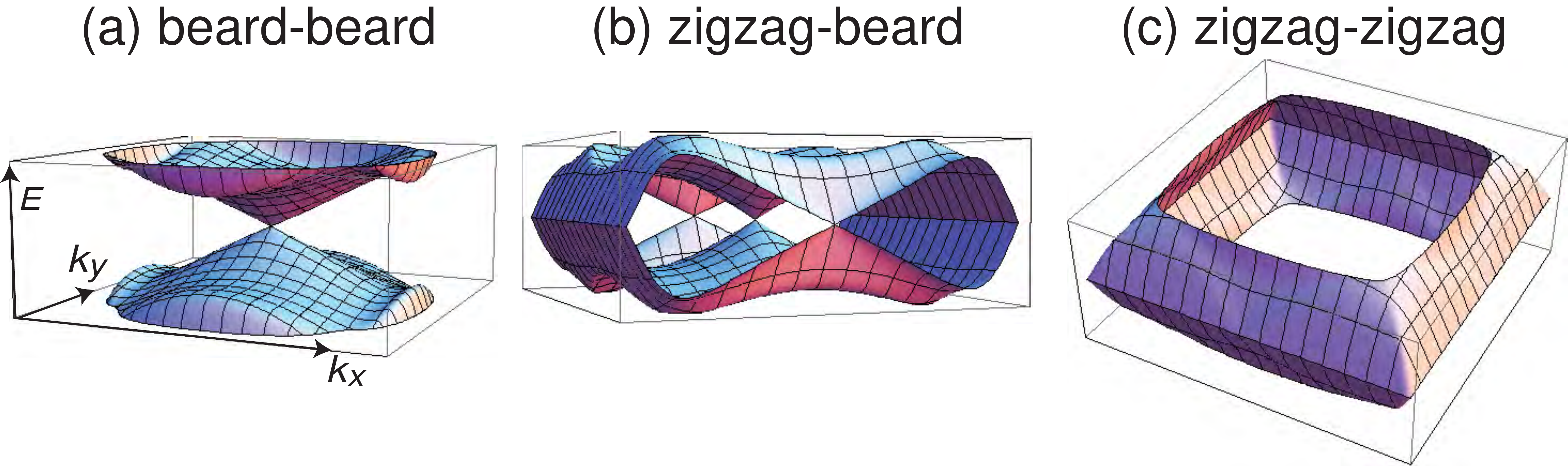}}
\caption{Band structure of a nanofilm with the SOI for (a) beard-beard edge,
(b) zigzag-beard edge and (c) zigzag-zigzag edge. One (two) Dirac cones
appear in the beard-beard (zigzag-beard) edge. We have set 
$\protect\lambda _{xy}=\protect\lambda _{z}=0.2t/2\protect\sqrt{3}$ with $t=t_{xy}=t_{z}$. }
\label{FigTEdge}
\end{figure}

\textit{Spin-orbit interaction:} It has been shown for the hyperhoneycomb
system\cite{Lee} that the system turns into a strong topological insulator
in the presence of the SOI. The Kane-Mele type spin-orbit interaction is
given by\cite{KaneMele,FuKaneMele,Lee}$H_{\lambda }=i\mathbf{v}_{ij}\cdot 
\mathbf{\sigma },$ with $\mathbf{v}_{ij}=v_{\text{KM}}
\frac{\mathbf{r}_{ik}\times \mathbf{r}_{kj}}{\left\vert \mathbf{r}_{ik}\times \mathbf{r}_{kj}\right\vert },
$where $v_{\text{KM}}$ is the strength of the spin-orbit
interaction. We assume $v_{\text{KM}}=\lambda _{z}$ for the $z$-direction
and $v_{\text{KM}}=\lambda _{xy}$ for the in-plane direction. We show how
the flat-band edge states change by introducing the spin-orbit interaction.
The resultant edge states are shown in Fig.\ref{FigTEdge}. The flat-band
edge states are bent by the SOI and turn into the topological edge states.
For example, the edge states are well described by the tetragonal-warped
Dirac cone for the beard-beard edges. On the other hand, the shapes of the
topological edge states are much different from the Dirac spectrum for the
zigzag-zigzag and zigzag-beard edges.

\textit{Effective model:} There are two bands near the Fermi energy for each
spin. It is possible to solve these eigenstates $\psi _{\pm }$ explicitly as 
$H_{2N}\psi _{\pm }=\pm (2t_{xy}-t_{z})\psi _{\pm }$ with $\psi _{\pm }=(\pm
1,-1,\mp 1,1,\cdots )$ when $N$ is even. We derive the effective 4-band
model with the SOI in order to describe the physics near the Fermi energy.
By evaluating $\psi _{\pm }^{\dagger }H\psi _{\pm }$, we obtain the
effective 4-band theory, 
\begin{equation}
H_{\text{eff}}=d_{1}\tau _{z}+d_{2}\tau _{y}+(-W+\frac{2N_{x}}{N}\Lambda
_{y}\sigma _{x}+\frac{2N_{y}}{N}\Lambda _{x}\sigma _{y})\tau _{x},
\end{equation}
where 
\begin{align}
d_{1}+id_{2}& =t_{z}e^{ik_{z}}-t_{xy}e^{-i\frac{k_{z}}{2}}\sum_{\alpha =x,y}
\frac{2N_{\alpha }}{N}\cos \frac{\sqrt{3}k_{\alpha }}{2}, \\
\Lambda _{\alpha }& =\lambda _{z}\sin \sqrt{3}k_{\alpha }+2\lambda _{xy}\sin 
\frac{\sqrt{3}k_{\alpha }}{2}\cos \frac{3k_{z}}{2},
\end{align}
with $\alpha =x,y$, and $W=V+\mathbf{V}_{\text{AF}}\cdot \mathbf{\sigma }$.
Here we have additionally included the staggered potential $V$ and the
antiferromagnetic staggered potential $\mathbf{V}_{\text{AF}}$ between the
two sublattices in the bipartite system [Fig.\ref{FigYJunc}].

\textit{The }$\mathbb{Z}_{2}$\textit{\ index:} We calculate the $\mathbb{Z}_{2}$ 
index for $V=\left\vert \mathbf{V}_{\text{AF}}\right\vert =0$. There
are the time-reversal symmetry and the inversion symmetry for $V=\left\vert 
\mathbf{V}_{\text{AF}}\right\vert =0$. The inversion symmetry operator is
given by $P=\tau _{z}$ with $PH_{\text{eff}}(\mathbf{k})P^{-1}=H(-\mathbf{k}) $. 
Then the $\mathbb{Z}_{2}$ index $\nu $ ($=0,1$) is given by the product
of the parity of $d_{1}$ at the 8 high-symmetry points $\Gamma _{i}$: 
$(\sqrt{3}k_{x},\sqrt{3}k_{y},k_{z})=$ $(0,0,0)$, $(2\pi ,0,0)$, $(0,2\pi ,0)$, 
$(\pi ,\pi ,0)$, $(0,0,2\pi )$, $(\pi ,0,2\pi )$, $(0,\pi ,\pi )$ and $(\pi ,\pi ,2\pi )$. 
The index $\nu $ is explicitly obtained as 
\begin{equation*}
(-1)^{\nu }=\prod_{i=1}^{8}\text{sgn}(d_{1}(\mathbf{k}=\Gamma _{i}))=\text{sgn}[t_{z}^{4}(t_{z}-2t_{xy})(t_{z}+2t_{xy})].
\end{equation*}
It follows that $\nu =1$ for $2|t_{xy}|>|t_{z}|$ and $\nu =0$ otherwise.
Consequently, the system becomes a strong insulator when the SOI is included
to the loop-nodal semimetal.

\begin{figure}[t]
\centerline{\includegraphics[width=0.5\textwidth]{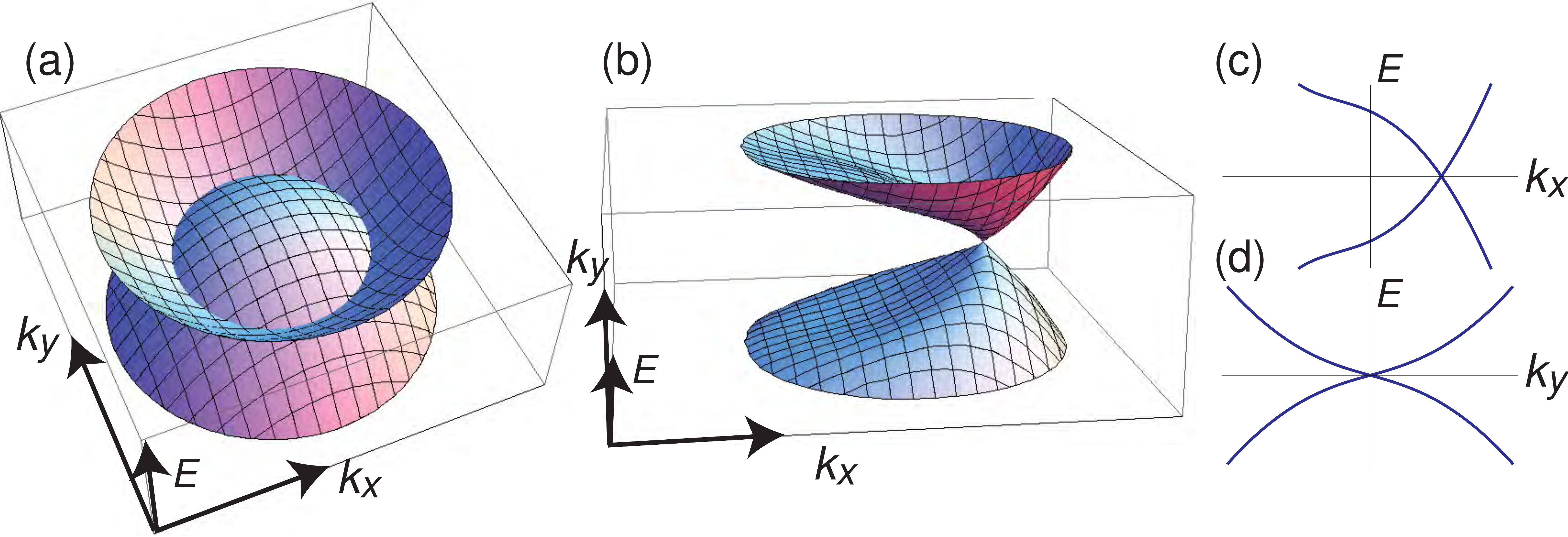}}
\caption{Bird's eye's view of (a) the Fermi loop, and (b) the point-nodal
semimetal with a Dirac cone. Band structure of the semimetal along (a) the $k_{x}$ axis 
and (b) the $k_{y}$ axis. Energy spectrum is linear at the Dirac
point. We have set $k_{z}=0$.}
\label{FigDirac}
\end{figure}

\textit{Dirac Semimetal:} We expand the Hamiltonian around the $\Gamma $
point as 
\begin{equation*}
H=[v^{2}(k_{x}^{2}+k_{y}^{2})-m]\tau _{z}-uk_{z}\tau _{y}+[-W+\Lambda
(k_{y}\sigma _{x}+k_{x}\sigma _{y})]\tau _{x},
\end{equation*}
where $v=\sqrt{3t_{xy}/8},u=t_{z}+t_{xy},m=2t_{xy}-t_{z}$ and 
$\Lambda =\sqrt{3}(\lambda _{z}+\lambda _{xy})$. The energy spectrum is easily
calculable, from which we find the following results: 
(i) When $V=\sqrt{m}\Lambda /v$ and $\mathbf{V}_{\text{AF}}=0$, 
the gap closes at a Fermi loop ($k_{x}^{2}+k_{y}^{2}=m/v^{2})$ as in Fig.\ref{FigDirac}(a); 
(ii) When $(V_{\text{AF}}^{x})^{2}+(V_{\text{AF}}^{y})^{2}=\Lambda ^{2}m/v^{2}$ and $V_{\text{AF}}^{z}=V=0$, 
the gap close at a Dirac point as in Fig.\ref{FigDirac}(b), (c) and (d).

By setting $\mathbf{V}_{\text{AF}}=\Lambda \sqrt{m}/v(\sin \theta ,\cos
\theta ,0)$, the position of the Dirac point reads $\mathbf{k}_{0}=\sqrt{m}/v(\sin \theta ,\cos \theta ,0)$. 
The Hamiltonian is linear around the Dirac
point,
\begin{equation}
H_{\text{eff}}=g_{x}(k_{x}-k_{x}^{0})+g_{y}(k_{y}-k_{y}^{0})+g_{z}k_{z},
\end{equation}
where $g_{x}=2\sqrt{m}v\sin \theta \tau _{z}+\Lambda \sigma _{y}\tau _{x}$, 
$g_{y}=2\sqrt{m}v\cos \theta \tau _{z}+\Lambda \sigma _{x}\tau _{x}$, 
and $g_{z}=-u\tau _{y}$ are constant $4\times 4$ matrices.

We have systematically constructed a wide class of the 3D honeycomb lattices
indexed by $[\alpha _{1}\alpha _{2}\cdots \alpha _{N}]$. All of them are
loop-nodal semimetals. Perfect flat bands will lead to the flat band
ferromagnetism when the Coulomb interaction is included. With the SOI, they
become strong topological insulators. It is intriguing that, with an
additional AF order, point-nodal semimetals with Dirac cones are generated
in the 3D space. Various 3D honeycomb lattices will be realized in the Li$_{2}$IrO$_{3}$ system 
since the building block of the 3D honeycomb lattice
is naturally realized by the octahedron of the Ir network\cite{Modic}. Our
results will open a new physics of the honeycomb system in the three
dimensions.

The authors is very much grateful to N. Nagaosa for many helpful
discussions on the subject. This work was supported in part by Grants-in-Aid for MEXT KAKENHI grant
number  25400317 and 15H05854.

\end{document}